\documentstyle[twocolumn,aps,prl,psfig]{revtex}
\begin{document}
\twocolumn[\hsize\textwidth\columnwidth\hsize\csname 
@twocolumnfalse\endcsname
\title{The Small Polaron Crossover: Role of Dimensionality}
\author{M. Capone}
\address{I.N.F.M. and
International School for Advanced Studies, SISSA-ISAS, \\
Trieste, Italy 34013}
\author{S. Ciuchi} 
\address{Dipartimento di Fisica, Universit\'{a} de L'Aquila,
via Vetoio, 67100 Coppito-L'Aquila, Italy and I.N.F.M., 
Unit\'a
de L'Aquila.}
\author{C. Grimaldi $^*$}
\address{I.N.F.M., Unit\'{a} di Roma 1,
Dipartimento di Fisica, Universit\'{a} di Roma ``La 
Sapienza".
P.le A. Moro 2, 00185 Roma, Italy}

\maketitle

\begin{abstract}
The crossover from quasi free electron to small polaron
in the Holstein model for a single electron
is studied by means of both exact and self-consistent calculations in
one dimension and on an infinite coordination lattice,
in order to understand the role of dimensionality in such a crossover.
We show that a small polaron ground-state occurs when both strong coupling
($\lambda>1$)
and multiphonon ($\alpha^2 >1$) conditions are fulfilled leading to
different relevant coupling constants ($\lambda$) in adiabatic
and ($\alpha^2$) anti adiabatic region of the parameters space. 
We also show that the self-consistent calculations obtained
by including the first electron-phonon vertex correction give accurate
results in a sizeable region of the phase diagram well separated
from the polaronic crossover. 
\end{abstract}

\vskip 2pc ] 

\narrowtext

\section{INTRODUCTION}

Recent optical measurements of the insulating parent 
compounds of the high-temperature superconductors 
show the presence of polaronic carriers\cite{htc}, and 
evidence for intermediate and strong lattice distortions   
has been given also for the 
colossal magnetoresistance manganites\cite{Mn} and Nickel 
compounds\cite{Ni}.
The recent observation of one-dimensional stripes in 
the high-temperature superconductors\cite{BISCO} and in manganites
suggests a comprehensive study of the role of dimensionality
in the polaronic crossover.  A detailed study of the small polaron crossover
is demanded also by the recent experimental results on 
manganites \cite{lanzara}.

The polaronic state is characterized by strong local electron-lattice 
correlation and is a non-perturbative phenomenon.
It therefore 
cannot be described by simple summation of the 
perturbative series such as the one which defines the 
Migdal-Eliashberg (ME) theory.
Here, we provide a detailed study of the 
crossover which occurs at intermediate electron-lattice 
couplings from quasi-free electron to small polaron ground state,
with a particular emphasis on the role of system dimensionality.

We consider the simple Holstein molecular-crystal Hamiltonian
for a single electron, which reads:

\begin{equation}
\label{holham}
{\cal H}=-t\sum_{\langle ij\rangle}
c^\dagger_i c_j +
g\sum_i
c^\dagger_i c_i\left( a_i+a^\dagger_i \right)
+\omega_0 \sum_i a^\dagger_i a_i
\end{equation}
where $c_i$ ($c_i^{\dagger}$) and $a_i$ ($a_i^{\dagger}$) are, 
respectively, the destruction 
(creation) operators for an electron and for a dispersionless phonon 
of frequency $\omega_0$ on site $i$. 
The hamiltonian (\ref{holham}) represents
a non-trivial many-body problem 
and it has been already studied in recent years 
by means of numerical 
\cite{marsiglio,deraedt,fehske,capone} 
and analytical \cite{alexandrov,lavorone,zhao} techniques.

Two dimensionless 
parameters are introduced to measure the strenght of electron-phonon
($el$-$ph$)
interaction: $\lambda=g^2/(D\omega_0)$ and $\alpha=g/\omega_0$,
where $D = 2td$ is the half-bandwidth
for the free electron and $d$ is the system dimensionality.

$\lambda$ is originally introduced in the weak 
coupling pertubation theory ($g/t\ll 1$) and 
is the coupling parameter of a ME approach in the case of one electron.
It can also be viewed as the ratio between the small polaron energy
$E_{\mbox{p}} = -g^2/\omega_0$ and the free electron energy 
$E_{\mbox{free}} = -D$.

The parameter $\alpha$ is the relevant coupling in the  {\it atomic
limit} ($t=0$). In this limit $\alpha$ measures the lattice 
displacement associated to the polaron and 
$\alpha^2$ is the average number of phonons bound to the 
electron.
According to the Lang-Firsov results
followed by the Holstein approximation, $\alpha$
also rules the reduction of the effective
hopping $t^{*} = t\exp{(-\alpha^2)}$ \cite{capone,ciuchi}.

Besides $\lambda$ and $\alpha$, the $el$-$ph$ system described by
eq.(\ref{holham}) is governed also by another dimensionless 
parameter: $\omega_0/t$. It measures
the degree of adiabaticy of the lattice motion (lattice kinetic 
energy $\simeq \omega_0$)
compared to the electron one (electron kinetic energy 
$\simeq t$).
In the adiabatic regime ($\omega_0/t \ll 
1$), $\lambda > 1$  is a condition sufficient to give a polaronic state
since the electron is bound to the slowly moving lattice giving
rise to a strong enhancement of effective mass.
In the antiadiabatic regime ($\omega_0/t \gg 1$) such a
picture is no longer true due to the fast lattice motion.
In this case, polaronic features such as strong local 
electron-lattice correlations arise only when the electron is
bound to a {\it large} number of phonons ($\alpha^2>1$).
To summarize, in both adiabatic and
antiadiabatic regimes, a polaronic state is formed when
{\it both} $\lambda > 1$ and $\alpha^2 > 1$ inequalities
are fulfilled \cite{capone}. This conclusion is in contrast
with ref.\cite{alexandrov} where it is argued that
 $\lambda > 1$ is the
only condition for small polaron formation. 

\begin{figure}
\protect
\vbox to 7.5cm{\vfill
\centerline{\psfig{figure=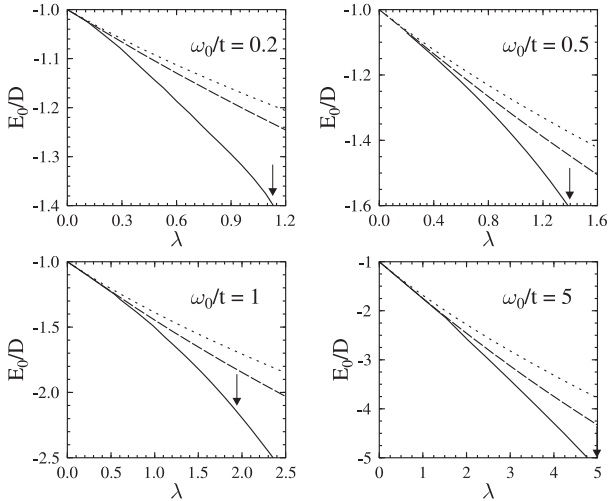,width=8cm}}
\vfill}
\caption{Ground state energy results in d=1. The exact 
diagonalization 
results are compared with the NCA (short dashed) and VCA 
(long dashed) calculations. The arrows mark $\lambda_c$
(see Fig.\protect\ref{fig-fase-d1}).}
\label{fig-energia-d1}
\end{figure}

The parameter $\omega_0/t$ influences also the dependence
of the behavior of the $el$-$ph$ coupled system on the system
dimensionality.
We shall show that in the antiadiabatic regime the 
small polaron formation does not depend on the system dimensionality.
On the other hand, dimensionality plays a crucial role in 
the adiabatic regime $\omega_0 \ll t$. This can be traced back 
to the adiabatic limit $\omega_0/t = 0$. In fact, in d=1 the ground state 
is localized for any finite value of $\lambda$ and 
a crossover occurs between {\it large} and 
{\it small} polaron at $\lambda\simeq 1$, whereas for $d \ge 2$ it has
been shown that a localization transition occurs at finite $\lambda$
from free electron to small polaron\cite{kabanov}.
The different adiabatic behaviors between 1d and 2d systems
could be relevant to describe the motion of polarons
as defects on top of 1d charge striped structures such as those
observed in cuprates\cite{BISCO} and manganites\cite{Mn}.

\section{RESULTS}

We study the relevance of
$\omega_0/t$ and of the lattice dimensionality d
by using two alternative exact calculations:
exact diagonalization of small one dimensional clusters (ED-1d) 
and dynamical mean field theory (DMFT-3d).
In the ED-1d approach,
the infinite phonon Hilbert space must be truncated to 
allow for a given maximum number of phonons per site 
$n_{\mbox{max}}$.
In order to properly describe the multiphonon regime (expecially
in the adiabatic regime where a large number of low energy 
phonons can be excited) we chose a cut-off of $n_{\mbox{max}} =20$. 
This 
high value forced us to restrict our analysis to a four-site cluster in the 
strong-coupling adiabatic regime. 
\begin{figure}
\protect
\vbox to 7.5cm{\vfill
\centerline{\psfig{figure=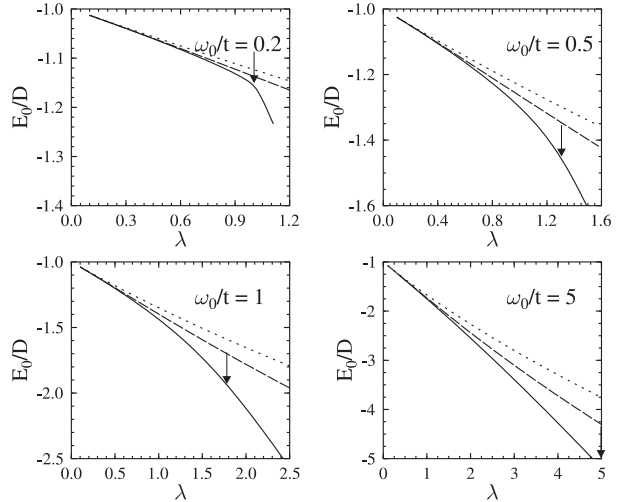,width=8cm}}
\vfill}
\caption{Ground state energy results for an infinite 
coordination
lattice. Comparison between
dynamical mean field (solid line), NCA (short dashed) and 
VCA (long dashed). The arrows mark $\lambda_c$
(see Fig.\protect\ref{fig-fase-d1}).} 
\label{fig-energia-d3}
\end{figure}
In the weak-coupling regime and for 
larger phonon frequencies
a lower value of $n_{\mbox{max}}$ is needed, allowing us to 
consider larger clusters up to ten-twelve sites.
We checked that finite-size effects do not affect the crossover coupling, 
since small-polaron formation is a local, high energy process.

The dynamical mean field theory approach
can be seen as the exact solution of the small polaron  problem
on an infinite coordination lattice\cite{lavorone}. 
The formulation of the DMFT requires the knowledge of the 
free particle DOS. A semi-circular DOS can mimic a  
three-dimensional case: in the following we will therefore refer to 
this approach to as  DMFT-3d. 

We calculate the exact ground state energy $E_0$
obtained by means of  ED-1d and DMFT-3d  and we compare the results with 
the  self-consistent non-crossing (NCA) 
and vertex corrected approximations (VCA).
These two approximations are defined by the self-consistent 
calculation of the electronic zero-temperature 
self-energy $\Sigma(k,\omega)$ given below:

\begin{eqnarray}
\label{selfen}
&\Sigma(k,\omega) \!= \! \frac{2\lambda\omega_0 
t}{N}\sum_{p}G(p,\omega-\omega_0) \times \nonumber \\ 
& \times\left[1\! +\!\frac{2\lambda\omega_0 t}{N}\sum_{q}G(q-
p+k,\omega-\omega_0)
G(q,\omega-2\omega_0)\right] ,
\end{eqnarray}
where $G(k,\omega)$ is the retarded fully renormalized  
single electron Green's function:

\begin{equation}
\label{green}
G(k,\omega)^{-1}=\omega-\epsilon_k-\Sigma(k,\omega)+
i\delta .
\end{equation}
which will be determined self-consistently. 
From eqs.(\ref{selfen},\ref{green}), the ground state
energy $E_0$ is given by the lowest energy solution of 
$\mbox{Re}\, G(k,E_0)^{-1}=0$.
The NCA approach amounts to compute $\Sigma$ by
retaining only the $1$ in the square brackets of eq. 
(\ref{selfen}). 
The VCA is instead given by the inclusion also of the second term 
in square brackets of eq.(\ref{selfen}) which represents
the first vertex correction.
This approach is formally similar to the approximation scheme 
used in the formulation of the non-adiabatic theory of superconductivity 
\cite{grimaldi} and a
comparison with exact results therefore provides also a test 
of reliability of such an approach for the one-electron case.

In fig.\ref{fig-energia-d1} we compare the ground-state energy 
$E_0$ obtained by ED-1d with the NCA and VCA results. 
The same quantities evaluated in the DMFT-3d case are shown 
in fig. \ref{fig-energia-d3}.
We have chosen the same half-bandwidth $D$ in both DMFT-3d and ED-1d.
In the adiabatic regime, the agreement of both approximations
with exact results strongly depends on the system 
dimensionality as a result of the different low-energy behaviour of the 
DOS . In fact, moving from $\omega_0/t=0.2$ to $\omega_0/t=0.5$, before
the crossover 
the agreement of the self-consistent calculations with the exact results is 
improved for the 1d case (fig. \ref{fig-energia-d1}) whereas it becomes 
poorer for the 3d one (fig. \ref{fig-energia-d3}).
However, the VCA approach represents a significative 
improvement with respect to the NCA for every 
system dimensionality and over a significant range of parameters.

As it is seen from the comparison
of fig. \ref{fig-energia-d1} and \ref{fig-energia-d3},
for large $\omega_0/t$  both approximate and exact results
tend to become independent of dimensionality. 
This can be understood by realizing that in this regime
the system can
be thought as a flat band ``atomic" system in interaction with 
high energy phonons.
It is also clear from figs. \ref{fig-energia-d1} 
and \ref{fig-energia-d3} that
both the self-consistent NCA and VCA calculations deviate
from the exact results when the crossover towards the small 
polaron regime is approached. 

\begin{figure}
\vbox to 13cm{\vfill
\centerline{\psfig{figure=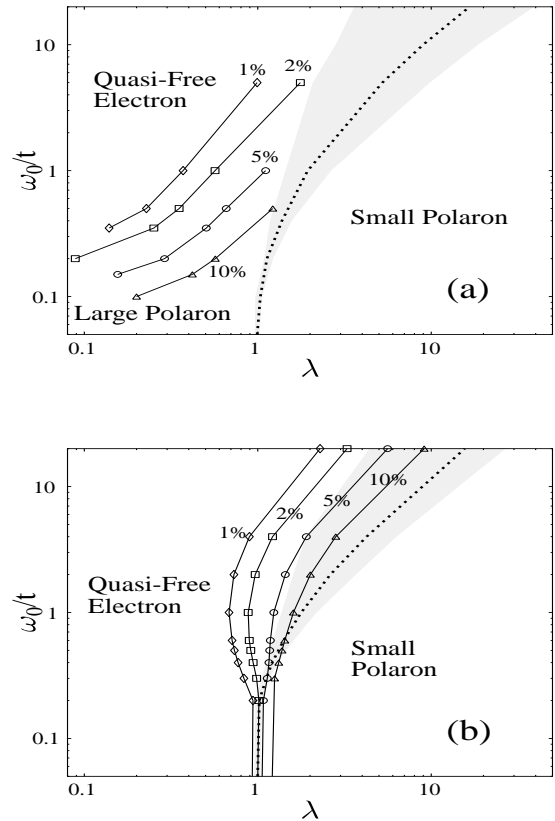,width=10cm}}
\vfill}
\caption{Phase diagram in the $\lambda$-$\omega_0/t$ plane 
for the one-dimensional (a)
and the infinite coordination lattice (b) Holstein model.
The dotted line is the polaron crossover value $\lambda_c$ 
and the width of the crossover is 
evidentiated by a shaded area. 
The isolines represents the relative difference 
between the exact and the VCA result for the ground state 
energy (see text).}
\label{fig-fase-d1}
\end{figure}

A complete comparison between 
the exact results and the VCA approach in the parameter
space $\lambda$-$\omega_0/t$ is shown in figs. \ref{fig-fase-d1}(a)-(b). 
We explicitly evaluated both in 1d (a) and 3d (b) the relative difference 
$\delta E_0  = 2|E_0^{\mbox{VCA}} - E_0^{\mbox{exact}}|/
|E_0^{\mbox{VCA}}+E_0^{\mbox{exact}}|$ 
where $E_0^{\mbox{exact}}$ and  $E_0^{\mbox{VCA}}$ are the 
ground-state energies evaluated by exact techniques and the VCA, 
respectively.
To analyze the region in the parameter space where the VCA agrees within 
a given accuracy with the exact results 
we report in figs. \ref{fig-fase-d1}(a)-(b) 
lines of constant $\delta E_0$.

As already mentioned, the agreement between self-consistent approximations 
and exact results is sensible to dimensionality.
For $d>2$, approaching the adiabatic 
limit and for small couplings the electron tends to be free.
For this reason self-consistent approximations work well.
On the contrary, in the adiabatic limit and for d=1 the ground state 
is a localized large polaron and self-consistent approximations fail to 
predict its energy.
In general,  VCA (and so NCA) works well outside the
polaron region whatever polarons are either small or large.
This can be seen directly from figs. \ref{fig-fase-d1}(a)-(b) 
where the critical 
coupling $\lambda_{\mbox{c}}$ of the crossover to small polaron  
is depicted as a dotted line. 
The critical coupling $\lambda_c$  is defined as the value at which
$d E_0/d g$ has maximum slope.
In the same figures, we provide
also an estimate of the width of the crossover (shaded areas)
obtained by looking at
the maximum slope of $|\partial^2 E_0/\partial g^2|$.
We checked that different criteria, like {\it e.g.} 
the effective mass enhancement\cite{lavorone}, 
provide the same qualitative results.

\section{CONCLUSIONS}

In conclusion, we have shown that the crossover towards
the small polaron state depends strongly on the adiabaticity
parameter $\omega_0/t$. In the antiadiabatic regime
the crossover is ruled by $\alpha^2$ and it is independent
of the system dimensionality. In the adiabatic regime
the relevant coupling is $\lambda$ and the crossover
occurs from large to small polaron in 1d, while in 3d the crossover
is from quasi free electrons to small polarons.
In the latter case self-consistent approximations work better than
in 1d systems.
We have also shown that self-consistent calculations
provide ground state energies which agree well with
exact results outside the small and large polaron region of the phase diagram
and that such an agreement is increased when vertex corrections are
taken into account. 

We thank M. Grilli, F. de Pasquale, D. Feinberg and 
L. Pietronero for stimulating
discussions. C. G. acknowledges the support of a I.N.F.M. 
PRA project.

\vskip 2pc
* Present address: \'Ecole Polytechnique F\'ed\'erale de
Lausanne, DMT-IPM, CH-1015 Lausanne, Switzerland.

\end{document}